\documentclass[conference]{IEEEtran}
\pdfoutput=1
\IEEEoverridecommandlockouts

\usepackage{cite}
\usepackage{amsmath,amssymb,amsfonts}
\usepackage{graphicx}
\usepackage{textcomp}
\usepackage{xcolor}
\def\BibTeX{{\rm B\kern-.05em{\sc i\kern-.025em b}\kern-.08em
    T\kern-.1667em\lower.7ex\hbox{E}\kern-.125emX}}

\usepackage{amsmath}
\usepackage{url}
\usepackage{hyperref}

\usepackage{enumerate}  
\usepackage{fancyhdr}
\usepackage{graphicx}
\usepackage{float}
\usepackage{caption}
\usepackage{tabto}
\usepackage{subcaption}
\usepackage{textgreek}
\usepackage{fixltx2e}
\usepackage{enumitem}

\usepackage{algorithm}
\usepackage[noend]{algorithmic}
\usepackage{comment}

\usepackage{soul}
\usepackage[utf8]{inputenc}

\usepackage{amsthm}
\usepackage{booktabs} %
\usepackage{tabularx}

\usepackage{xcolor}
\newtheorem{problem}{Problem}

\begin{document}

\title{Global Prediction of COVID-19 Variant Emergence
Using Dynamics-Informed Graph Neural Networks}

\author{
    \IEEEauthorblockN{Majd Al Aawar}
    \IEEEauthorblockA{
        \textit{Ming Hsieh Department of ECE} \\
        \textit{University of Southern California, USA}\\
        malaawar@usc.edu
    }
    \and
    \IEEEauthorblockN{Srikar Mutnuri}
    \IEEEauthorblockA{
        \textit{Department of Computer Science} \\
        \textit{University of Virginia, USA}\\
        nmp8rj@virginia.edu
    }
    \and
    \IEEEauthorblockN{Mansooreh Montazerin}
    \IEEEauthorblockA{
        \textit{Ming Hsieh Department of ECE} \\
        \textit{University of Southern California, USA}\\
        mmontaze@usc.edu
    }
    \and
    \IEEEauthorblockN{Ajitesh Srivastava}
    \IEEEauthorblockA{ 
        \centerline{Ming Hsieh Department of ECE} \\
        \centerline{University of Southern California, USA}\\
        ajiteshs@usc.edu
    }
}

\maketitle
\begin{abstract}
During the COVID-19 pandemic, a major driver of new surges has been the emergence of new variants. When a new variant emerges in one or more countries, other nations monitor its spread in preparation for its potential arrival. The impact of the new variant and the timings of epidemic peaks in a country highly depend on when the variant arrives. The current methods for predicting the spread of new variants rely on statistical modeling, however, these methods work only when the new variant has already arrived in the region of interest and has a significant prevalence. Can we predict when a variant existing elsewhere will arrive in a given region? To address this question, we propose a variant-dynamics-informed Graph Neural Network (GNN) approach. First, we derive the dynamics of variant prevalence across pairs of regions (countries) that apply to a large class of epidemic models. The dynamics motivate the introduction of certain features in the GNN. We demonstrate that our proposed dynamics-informed GNN outperforms all the baselines, including the currently pervasive framework of Physics-Informed Neural Networks (PINNs). To advance research in this area, we introduce a benchmarking tool to assess a user-defined model's prediction performance across 87 countries and 36 variants.
\end{abstract}

\section{Introduction}

The COVID-19 pandemic presented an unprecedented global health crisis, severely affecting millions of people worldwide and demanding swift and effective responses from governments and healthcare providers~\cite{WHO}. As the pandemic progressed, multiple variants of COVID-19 emerged, each possessing genetic mutations that can significantly impact transmissibility, virulence, and even vaccine efficacy. During much of the COVID-19 epidemic, the surge in cases and severe outcomes (hospitalizations and deaths) have been driven by the emergence of new variants~\cite{Wiemken2023}. Consequently, monitoring the emergence and spread of these variants is crucial for devising appropriate public health measures and optimizing containment strategies~\cite{WHO2}. A critical factor driving the surge in a given region is the arrival time of the new variant~\cite{Markov2023}. We can observe how a new variant, that has not appeared in region $A$ yet, spreads in region $B$. We can study the spread in the region $B$ to understand the properties of this new variant. If it is a highly transmissible or immune-evading variant~\cite{lambrou2022genomic}, we can expect it to spread in the region $A$ eventually. However, precisely when it would happen in region $A$ remains unknown. Making a good prediction of arrival time would lead to more effective preparation and resource management.

We focus on the problem of predicting the arrival of a new variant in a given region provided that it has appeared somewhere else. Due to testing delays and the fact that not all cases are genomically analyzed, it is difficult to pin down when a variant arrives. Therefore, we consider measuring the delay to reach a certain proportion of prevalence.

Figure~\ref{fig:intro-covariants} shows the proportions of different variants over time in the United Kingdom and Sweden. We can see that there is often a clear delay (see 20.Alpha.V1 and 21.J.Delta) in arrival of the variants in Sweden.
Furthermore, a variant may start to spread, but quickly get dominated by another variant before it reaches a significant proportion of the circulating cases (e.g., 21.A.Delta in Figure~\ref{fig:intro-covariants}).
Therefore, we reformulate the problem as the following:

\begin{figure}[!ht]
    \centering
    \includegraphics[width=0.80\linewidth]{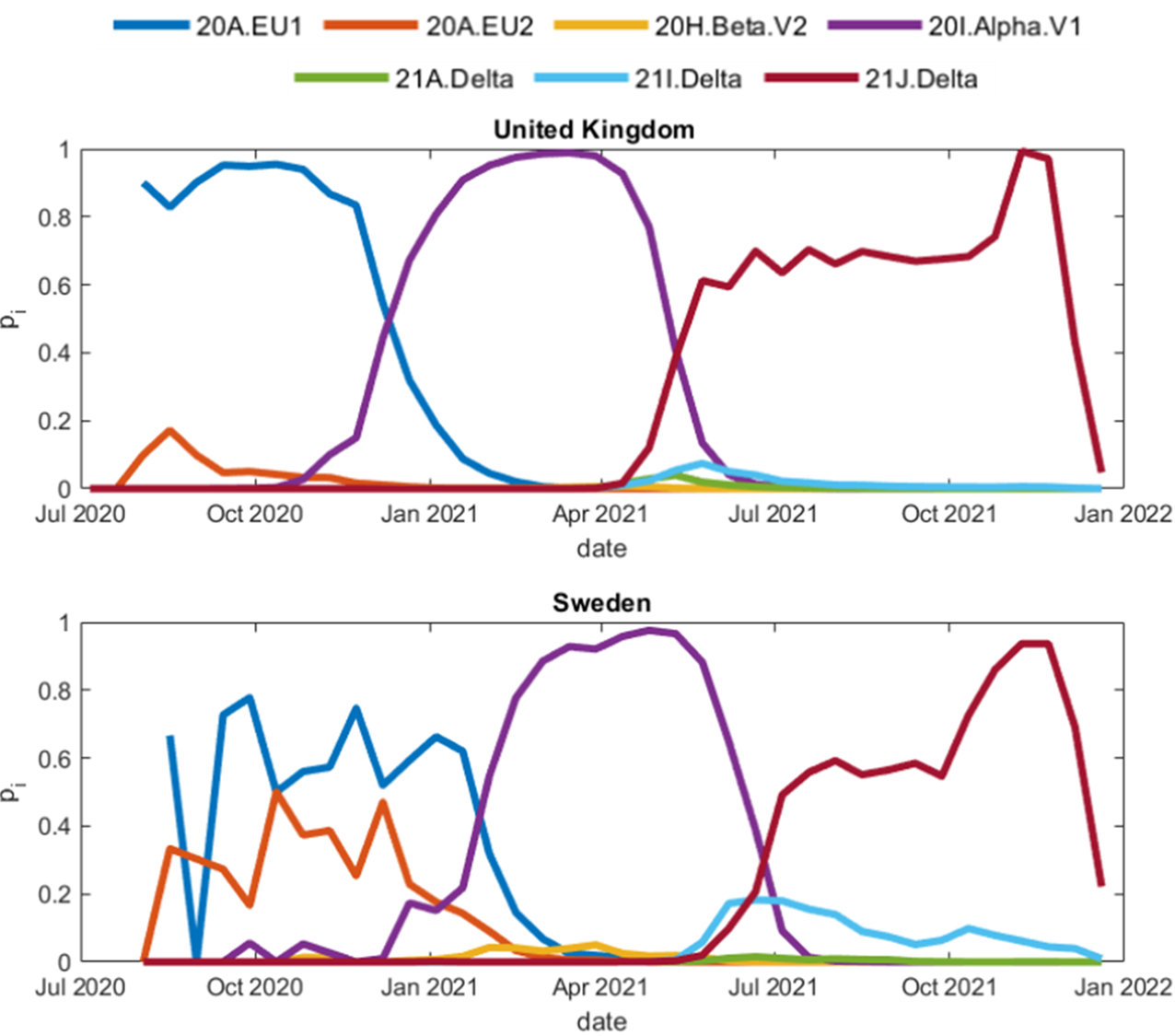}
    \caption{\small Plots showing the prevalence of a few different variants of COVID-19 over time in the UK and Sweden. The beginning of the spread of new variants can differ across countries by several weeks. This can be seen when examining ``20I.Alpha.V1" and ``21J.Delta" variants in the provided plots, showing their appearance in the UK several weeks before reaching Sweden.}
    \label{fig:intro-covariants}
\end{figure}

\begin{problem}[Emergence Delay Prediction]
    Given the prevalence (proportion) of a variant in regions $A_1, A_2, \dots$, predict \textbf{when} the variant will reach a proportion of $\theta$ in region $B$, provided that this variant has not yet reached region $B$.
\end{problem}

Here, the term ``proportion" refers to the fraction of new cases created by the variant under consideration out of total new cases. 
Note that this is a regression problem -- we seek to predict a quantity, which is the delay between the current date and the date of reaching the desired threshold. Further, we wish to perform this prediction for a given region before any sample of the variant of interest is observed in that region.

Prior works have focused on the problem of predicting the prevalence of a variant when it has already arrived in the region of interest~\cite{sun2020forecasting, hu2020early}. Specifically, logistic regression turns out to provide a good estimate of the prevalence over time~\cite{lr01, lr02}. However, these techniques require the variant to have a non-zero proportion already in the region of interest. To the best of our knowledge, no prior work exists predicting when the delay reaches a certain prevalence even when there is zero prevalence in the region.

When a new variant emerges with evolutionarily favorable properties, it can only be transferred to a different region through a host (in the case of COVID-19 -- a human). This encourages the idea of using an underlying network of mobility to address the proposed problem. Since reaching a certain prevalence may depend on other currently circulating variants, their dynamics (how fast one variant can spread over others) also play a role. Therefore, we propose a variant dynamics-informed Graph Neural Network (GNN) that utilizes a network of mobility and features inspired by variant dynamics to solve the proposed problem. This is different than typical Physics-Informed Neural Networks (PINNs) where dynamics act as a regularization for the loss~\cite{raissi2019physics}. Here, we show that our approach of constructing appropriate features results in lower errors compared to the PINNs approach.

Our key contributions represent a novel effort in addressing the challenge of predicting the emergence of COVID-19 variants at a global level, and in doing so establishing a new benchmark for evaluating Delay Prediction on the ``CoVariants" dataset~\cite{covariants}. We expect that this benchmark will help build the capacity to predict arrival times of the future variants of COVID-19 and other outbreaks.
More specifically, our contributions are as follows: 

\begin{enumerate}
    \item We develop novel adaptations of GNNs that account for complex inter-dependencies between countries using GNNs while incorporating variant delay dynamics at the node level. To the best of our knowledge, we are the first to derive these delay dynamics for variants.
    Experiments demonstrate that our approach leads to superior results compared to several Machine Learning (ML) methods, including the currently pervasive framework of PINNs that integrate dynamics in the loss function.
    \item We make our evaluation pipeline publicly available so that it can be used to evaluate any user-defined PyTorch model\footnote{https://github.com/itssmutnuri/gnnvariants}. 
    This will advance public health research and enable public health experts and policymakers to leverage this benchmark pipeline for improvement of their models and consequently, for enhancing global health outcomes.
\end{enumerate}

\section{Related Work and Background}

\subsection{Variants and Variant Dynamics}

In the context of infectious diseases, the term ``variant" refers to a version of a virus with some changes in its genetic material, known as genome~\cite{CDC_class}. These changes happen through genetic mutations and can affect the virus's characteristics, like how easily it spreads, how severe the illness it causes is, and whether it can evade the immune system or not. Variants can be categorized in multiple ways based on genetic differences, and the significance of these differences can vary. Some categorizations focus on a common ancestor, while others may specifically highlight mutations in key regions of the virus's genome~\cite{CDC_class}. In our study, we rely on the categorizations provided by the CoVariants dataset \cite{covariants}.

\begin{figure}[!ht]
\centering
\includegraphics[width=\columnwidth]{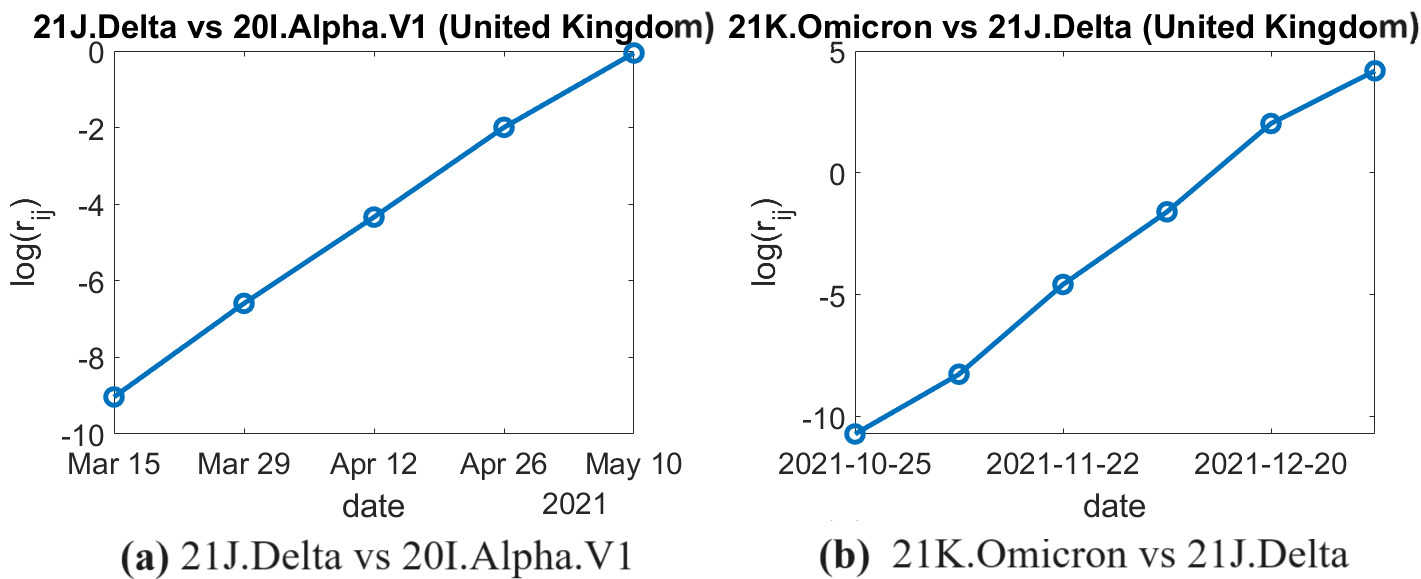}
\caption{\small Sample semi-log plots of variant proportions from data in the CoVariants dataset.}
\label{fig:Semilog}
\end{figure}

Furthermore, the proportions of different variants can be analyzed over time. Some studies have observed that these proportions follow a straight line when plotted on a semi-logarithmic scale. They leverage this fact to understand the dynamics of variant prevalence in populations \cite{Beesley2023}. We also observe this in our data as shown in Figure \ref{fig:Semilog}. However, to the best of our knowledge, existing studies do not derive any dynamics capable of predicting delays in the emergence of variants between multiple regions/countries.

\subsection{Graph Neural Networks and Epidemics}

Many efforts were made to develop forecasting models capable of predicting the progression and future trends of the COVID-19 pandemic \cite{US,EU,DU}. However, these forecasting approaches have primarily focused on predicting the spread of the virus rather than predicting delays in variant emergence. Predicting the timing of variant-specific outbreaks can significantly enhance public health preparedness and response strategies, allowing for timely interventions such as targeted vaccination campaigns and tailored public health measures \cite{WHO2}.

Many Deep Learning (DL) techniques have been explored for this purpose, including GNNs which help capture the graph effects of the spread of infectious diseases over time across regions \cite{scarselli2008graph, GTNN, STAN, transfergraph, pinnganesan2021spatio}. Other methods like PINNs ~\cite{pinnseo2019physics, pinnseo2019differentiable, pinnraissi2019physics}, agent-based models \cite{chopra2022differentiable, chopra2021deepabm}, and a combination of spatio-temporal networks with location-aware features \cite{pinndeng2019graph}, including mobility data \cite{pinnkapoor2020examining} have also been used.

Instead of only predicting virus spread, as others have done, we aim to predict when a new variant will emerge in a region given that it has emerged elsewhere. This involves deriving dynamics of variant spread and integrating these insights into a GNN. Notably, our strategy revolves around crafting relevant features rather than altering the loss function, providing a distinctive and effective approach.
 
\section{Methodology}

\subsection{Variant Dynamics} \label{ssec:vdynamics}

First, we develop an understanding of the dynamics that play a role in the spread of a new variant and how we can utilize them to make informed predictions. We start by extracting each variant's \textit{global growth rate} from observed data. This value provides an indication of the dominance of the variants over other competing ones. Subsequently, we construct linear models based on the dynamics of infectious diseases and the derived growth rates to predict delays between regions, establishing this \textit{Dynamics-based Model} as our baseline for \textit{Delay Prediction}. 
Finally, we consider different approaches in which we can incorporate some disease dynamics using GNNs that are chosen for their ability to intuitively model spatial dynamics, with the goal of enhancing the prediction accuracy of our model. 

Throughout the text, we use the terms ``prevalence" and ``prevalence ratio", where the ``prevalence", $p$, is the proportion of cases of a given variant, and the ``prevalence ratio" is defined as $r_{ij} = \frac{p_i}{p_j}$,
where $p_i$ is the prevalence of the variant $i$, and $p_j$ is the prevalence of the most prevalent variant, $j$, that is not $i$. In our experiments, a variant is considered dominant if it reaches an $r_{ij}$ value of at least $\frac{1}{3}$. We define this value as the \textit{dominance threshold}, $\theta$.

\subsection{Variant Growth Rates} \label{sec:growthrate} 

We first derive the dynamics of two competing variants in one region. While logistic regression has been widely used to fit these dynamics, for completeness, we present a derivation. We start by using a model of infectious diseases that generalizes a large class of models~\cite{SIkJalpha}:
\begin{align}
\Delta I(t) &\approx \beta \kappa(t) \bar{S}(t) \sum_{\tau=0}^{\infty} \bar{\alpha}(\tau) \Delta I(t - \tau)  \nonumber \\
&\approx \beta \bar{S}(t) \Delta I(t - b)(1 + \epsilon_t)
\end{align}

where $\Delta I(t)$ is the new infections at time $t$, $\bar{\alpha}(\tau)$ are transmission parameters, $\bar{S}(t)$ is the fraction of susceptible population, and $\epsilon_t$ is some small number. The number of contacts at time $t$ is given by $\kappa(t)$. The approximation follows from Taylor expansion, and using the fact that $\Delta I(t)$ is a smooth function. The constant $b$ happens to be the mean serial interval -- the average time between two successive infections. Now, given two variants $i$ and $j$, assuming full cross-immunity between all variants, i.e. $\bar{S}_i(t) =\bar{S}_j(t) = \bar{S}(t)$, the prevalence ratio is approximated as:
\begin{align}
r_{ij}(t) &= \frac{\Delta I_i(t)}{\Delta I_j(t)} \approx \frac{\beta_i}{\beta_j} \frac{\kappa(t) \bar{S}(t)\Delta I_i(t - b_i)}{\kappa(t)\bar{S}(t)\Delta I_j(t - b_j)} \nonumber \\ 
&\approx \frac{\beta_i}{\beta_j} \frac{\Delta I_i(t - b)}{\Delta I_j(t - b)} = \beta_{ij} r_{ij}(t - b)
\end{align}

The above further assumes that the mean serial intervals of the two variants are close, i.e., $b_i = b_j = b$.  Note that we assumed full cross-immunity which is true for a number of variants before the Omicron sub-variants~\cite{Chen2023}. If there is only a partial cross immunity,  $\bar{S}_i(t) \neq \bar{S}_j(t)$. However, in a short interval, the new immunity (reduction is susceptibility) created by the variants is small. As a result, it can be shown that   $\bar{S}_i(t) \approx \beta_{ij}' \bar{S}_j(t)$ for some constant $\beta_{ij}'$ which can be absorbed in the parameter $\beta_{ij}$ above.
Solving the above recurrence relation, we get:
\begin{equation*}
    r_{ij}(t) = (\beta_{ij})^{t/b} r_{ij}(0)
\end{equation*}

Taking the logarithm reduces the equation to: 

\begin{equation}
\label{eq:straight_line}
\ln(r_{ij}(t)) = \frac{t}{b} \ln(\beta_{ij}) + \ln(r_{ij}(0)) \\
= S_{ij}t + C
\end{equation}

Here, constant $b$ is the mean serial interval – the average time between two successive infections, and $\beta_{ij}$ is the ratio of transmission rates between the two competing variants. Parameter $S_{ij}$ can be referred to as the relative growth rate of variant $i$ over variant $j$~\cite{SIkJalpha}. 

Now, consider a scenario in two regions $A$ and $B$ --- $A$ has the variant $i$ emerging over the previously dominant variant $j$; but variant $j$ never reached region $B$ where variant $k$ is dominant. This makes it difficult to assess the potential impact of variant $i$ on region $B$ when we only know $S_{ij}$. To deal with this challenge, we note that the parameter $S{ij}$ should ideally be independent of the region (this may not hold due to the simplifying assumptions above, so we actually have a \textbf{region-specific relative growth rate} $S^{(p)}_{ij}$ for region $p$). Therefore, we can attempt to find the growth-rate of any variant $i$ with a fixed variant $0$, specifically, the original COVID-19 variant. For ease of notation, we represent the growth advantage of variant $i$ with respect variant $0$ as $S_{i}$, with $S_{0} = 0$, and refer to these values as \textbf{global growth rates}. Ideally, $S_{ij} = S_i - S_j$, irrespective of the region where the $S_{ij}$ is estimated.

To estimate these the global growth rates for all the variants  that exist by the time $t$ over all regions $p$,  we can set up the following system of linear equations: 
\begin{equation}
\begin{aligned}
S_i - S_j = S^{(p)}_{ij} + \epsilon_{ijp} , \forall i,j,p,
\end{aligned}
\end{equation}

where $S^{(p)}_{ij}$ is the growth advantage of variant $i$ over variant $j$ in region $p$ and $\epsilon_{ijp}$ are error terms. The solution is given by minimizing the sum of squares of $\epsilon_{ijp}$.

We, then, define an objective function and solve for the $S_{i}$ value as
\begin{equation}
\sum_{i, j, p} \epsilon^2_{ijp}\,.
\end{equation}
Finally, based on the global growth rates, we define \textbf{global relative growth rates} $S_{ij}$ as $S_i - S_j$.

These growth rates are used in our proposed methods for \textit{Delay Prediction}. 

\subsection{Dynamics-based Model} \label{ssec:baseline} 

Consider three variants $i$, $j$, and $k$ where $i$ is an emerging variant. We assume that either of the following holds -- (i) Variants $j$ and $k$, respectively in regions $X$ and $Y$, are the variants that constitute most of the other infections; or (ii) All variants other than $i$ in the respective regions $X$ and $Y$ have similar enough global growth rates that can be grouped into variants $j$ and $k$. Suppose variant $i$ appears in region $X$ first, followed by region $Y$. Then we have:
\begin{equation}
\label{eq:1}
\ln \left(\frac{{p^{Y}_i(t)}}{{p^{Y}_j(t)}}\right) = S_{ij}(t-t_0) + C_Y'
\end{equation}
\begin{equation}
\label{eq:2}
\ln \left(\frac{{p^{X}_i(t)}}{{p^{X}_k(t)}}\right) = S_{ik}(t-t_0) + C_X'
\end{equation}
For a new variant, initially, the number of infections increase because of the new infections coming from a different region (importations). After some point, when transmission starts to happen within region (community transmission), the imported infections become negligible in comparison, and this is the point after which above equations are valid.

Suppose we select $t_0$ as the time at which $p^{X}_i(t_0)$ is large enough such that the community transmission dictates the dynamics in the region $Y$ rather than importations and so, both Equations \ref{eq:1} \& \ref{eq:2} hold for $t \geq t_0 $. Suppose that the threshold for prevalence in region $Y$ for community transmission to dominate is $\theta_Y$ and that for region $X$ is  $\theta_X$. We assume that $\theta_X$ and $\theta_Y$ are independent of the variant. Then, by setting $t=t_0$ in equations \ref{eq:1} \& \ref{eq:2}, we get:
\begin{align}
\label{eq:0}
\ln \left(\frac{{\theta_Y}}{{1 - \theta_Y}}\right) = C_Y' \quad \text{and} \quad
\ln \left(\frac{{\theta_X}}{{1 - \theta_X}}\right) = C_X'
\end{align}
Therefore, both $C_Y'$ \&  $C_X'$ are variant independent. Let $t^{\theta}_Y$ and $t^{\theta}_X$ be the times at which variant $i$ reach target prevalence of $\theta$ in regions $Y$ and $X$, respectively. We want to find $\tau = t^{\theta}_Y - t^{\theta}_X$. From equations \ref{eq:1} and \ref{eq:2}:
\begin{equation}
\label{eq:3}
\begin{aligned}
\ln \left(\frac{{\theta}}{{1 - \theta}}\right) = S_{ij}(t^{\theta}_Y - t_0) + C_Y' = S_{ik}(t^{\theta}_X - t_0) + C_X' \\
\implies S_{ij} \tau = (S_{ij} - S_{ik})(t^{\theta}_X - t_0) + C_X' - C_Y'
\end{aligned}
\end{equation}
Also from equation \ref{eq:2}:
\begin{equation}
\begin{gathered}
t^{\theta}_X - t_0 = \frac{1}{S_{ik}}\left(\ln \left(\frac{{\theta}}{{1 - \theta}}\right) - C_X'\right). \\
\end{gathered}
\end{equation}
Plugging this value into equation \ref{eq:3}, we get:
\begin{equation}
\label{eq:final}
\begin{gathered}
S_{ij} \tau = \frac{(S_{ij} - S_{ik})}{S_{ik}}\left(\ln \left(\frac{{\theta}}{{1 - \theta}}\right) - C_X' \right) + C_X' - C_Y' \\
\implies \tau = \frac{S_{kj}}{S_{ij}S_{ik}}\ln \left(\frac{{\theta}}{{1 - \theta}}\right) - \frac{S_{kj}}{S_{ij}S_{ik}}C_X' + \frac{C_X' - C_Y'}{S_{ij}}
\end{gathered}
\end{equation}
Here $\theta$ is a given prevalence, $S_{ij}$, $S_{ik}$, and $S_{kj}$ are global relative growth rates estimated as shown in Section~\ref{sec:growthrate}. $C'_X$ and $C'_Y$ are unknowns.
Therefore for each pair of regions $X$ and $Y$, to identify the delay in reaching a given prevalence $\theta$, we can build a linear model $Y = W^T Z$ 
\begin{equation}
\label{eq:LM}
\begin{gathered} 
    Y = \tau(i,j,k,\theta) - \frac{S_{kj}}{S_{ij}S_{ik}}\ln \left(\frac{{\theta}}{{1 - \theta}}\right) \\
    Z = \begin{bmatrix}
    - \frac{S_{kj}}{S_{ij}S_{ik}} \\
    \frac{1}{S_{ij}}
    \end{bmatrix}
    W = \begin{bmatrix}
    C_X'\\
    C_X' - C_Y'
    \end{bmatrix}
\end{gathered}
\end{equation}
We then use the calculated weights from our linear models in equation \ref{eq:LM} and plug them into equation \ref{eq:final} to find the $\tau$ value between two regions. This is a pairwise linear model that estimates the delay between two regions rather than providing a delay from the current date.  Algorithm \ref{alg:variant_arrival} details how we compute the date at which a variant arrives in a region $Y$. We calculate the median of the outputs from all pairwise models from regions $X_q$ to region $Y$. We evaluate the performance of all the models $(X_q, Y)$ until the current date and select the top three performing source regions among ${X_1, X_2, \dots}$. Notably, best performance here refers to the lowest error in predicting the delay. We identify three source regions that produced the lowest error in predicting the delays at $t-1$. Then, we take the median of the predictions produced by these selected models at time $t$ which represents the delay at time $t$ for the new variant to appear in region $Y$.

\begin{algorithm}
  \caption{Dynamics-based Arrival Date Computation}\label{alg:variant_arrival}
  \begin{algorithmic}[1]
    \FOR{$X_q$ in $\{X_1, X_2, \dots\}_t$}
      \STATE Calculate and store delay predictions for $(X_q, Y)$ until the current date, $t$.
    \ENDFOR
    \IF{$t \neq 0$}
      \STATE Select top 3 $X_q$ with the best performance at $t-1$
    \ELSE 
        \STATE Select all source regions $\{X_1, X_2, \dots\}_t$
    \ENDIF
    \STATE Find the median of delay predictions for selected models
    \STATE \textbf{return} Median delay as the estimated arrival date for the variant in region $Y$ at time $t$  
  \end{algorithmic}
\end{algorithm}

\subsection{Dynamics-Informed GNN}

\textbf{Key Idea:} We observe that global growth rates $S_i$ play a crucial role in the dynamics. Furthermore, based on Equation \ref{eq:straight_line}, under some assumptions, the logarithm of the prevalence ratio $r_{ij}$ grows linearly with time. We hypothesize that using $\ln(r_{ij})$ and $S$ as features simplify the underlying patterns to be learned by an ML algorithm. An ML algorithm may also complement any violations of the assumptions made in the \textit{dynamics-based model}. 

We propose a GNN-based network for this problem as GNNs are adept at capturing spatial relationships, making them well-suited for problems where geographical proximity between regions, such as countries, plays a crucial role. Additionally, the problem induces a natural graph structure, where countries can be represented as nodes with edges denoting some relationship between the countries. Furthermore, we explore different techniques to incorporate disease dynamics into the GNN in an effort to capture the complex interactions and patterns associated with the spread of infectious diseases.

\begin{figure*}[!ht]
\centering
 \includegraphics[width=0.75\textwidth]{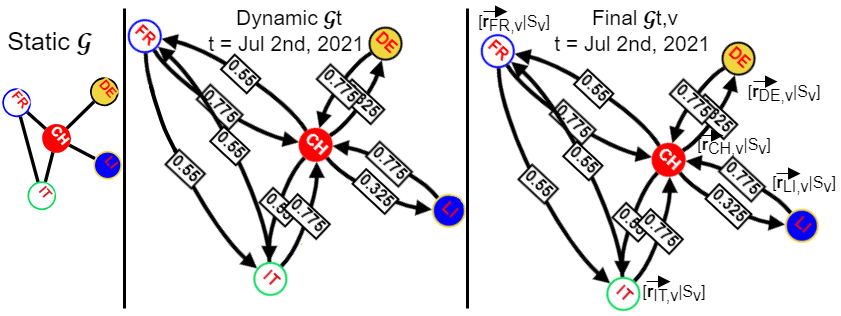}
 \caption{Graph creation process on sample subgraph $\mathcal{G}$: (i) first we construct our nodes and edges based on country adjacency (ii) Next, we account for temporal variations in the relations between countries, i.e. the edges,to get a $\mathcal{G}_t$ (iii) Finally, we find the variant specific features to get our sample graph $\mathcal{G}_{t,i}$}
 \label{fig:Graph Construction}
\end{figure*}

\subsubsection{Graph Construction} \label{sec:graph}

The graph construction is illustrated in Figure \ref{fig:Graph Construction}. First, we represent each country as a node in the graph, and edges are established to represent relationships between the countries. There are 87 countries in which at least one variant appears and becomes dominant in the CoVariants dataset \cite{covariants}. The dataset provides the state of variants and mutations of interest for COVID-19 from August 2020 through October 2023. It has a bi-weekly resolution for 36 variants starting from 20A.EU1 to 23F.Omicron. While not shown in Figure \ref{fig:Graph Construction}, self-loops are included to account for internal transmissions. 

Next, we consider the temporal aspect of the problem. The graph may evolve over time to capture changing relationships or influences among countries. This is introduced into our graph as dynamic edge weights based on border control data in the ``OxCGRT" dataset \cite{hale2021global}. This dataset is from January 2020 to January 2023 measuring the variations in government responses using their COVID-19 Government Response Stringency Index, which is a simple additive score of nine indicators ranging from school closures to vaccination policies. Of these indicators, we picked international travel closure controls since they best capture the emergence and cross-border transmission of variants. These values are encoded in an ordinal scale as follows: 0 - no measures, 1 - screening, 2 - quarantine on high-risk regions, 3 - ban on high-risk regions, 4 - total border closure. The recorded values were then scaled down to a range between 0.1 and 1, with 1 indicating no border restrictions and 0.1 representing complete border closure. Note that 0.1 was chosen over 0, acknowledging the possibility of some mobility between connecting countries due to necessary trade or border crossings via open land routes. Given that our CoVariants data extends beyond January 2023, we extend OxCGRT by assuming that all border restrictions have remained unchanged since Jan 8, 2023.

To obtain the full graph, we create multiple copies of the above graph representing regions and inter-connectivity -- one per variant at a given timestamp $t$. This approach allows us to leverage variant-specific dynamics as our node features. At $t$, each node $c$ has features consisting of a time series of the logarithm of prevalence ratios $r_{ij}$ of a present variant $i$ and the most prevalent variant $j \neq i$, spanning $T$ time steps. This results in a vector $\vec{r}_{c,i} = \left[\ln\left(r_{ij}^{t-T}\right), \ln\left(r_{ij}^{t-T+1}\right), \dots, \ln\left(r_{ij}^{t}\right)\right]$. The corresponding $S_i$ value of the variant $i$ is then concatenated to form each node's final feature vector $[\vec{r}_{c,i}|S_i]$.

\subsubsection{Feature Augmented Graph Convolutional Network (FA-GCN)}

\begin{figure*}[!ht]
\centering
 \includegraphics[width=\linewidth]{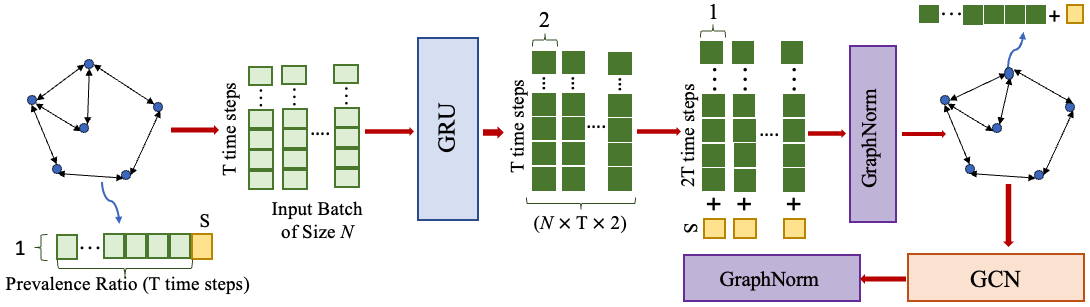}
 \caption{\small FA-GCN Architecture. The prevalence ratio for each time step is encoded into a 2D latent feature vector by the GRU. Then, the feature vectors are flattened and the growth rate is concatenated to them. Our batch size of $N$ = \# of countries.}
 \label{fig:FAGCN}
\end{figure*}

To extract both temporal and geographical information of the data, we propose a hybrid architecture summarized in Figure~\ref{fig:FAGCN}. In this model, we employ a simple Gated Recurrent Unit (GRU)~\cite{cho2014learning} to extract temporal information of the variant transmissions in $T$ time steps. We temporarily remove the growth rate $S_i$ (which is not time-dependent) from node features and embed the 1-dimensional (1D) time-dependent features (prevalence ratios) into a 2D latent feature vector. Note that in this configuration, we embed the features separately for each country, treating the nodes as distinct input samples grouped into a single batch, resulting in a 3D latent feature vector. The 3D latent feature vector is then flattened to obtain a final 2D embedding, which serves as input for a one-layer Graph Convolutional Network (GCN). The Leaky ReLU activation function introduces non-linearity after the GCN layer. Additionally, two GraphNorm\cite{GraphNorm} layers are applied after the GRU and GCN layers for normalization, and dropout is incorporated after the GCN layer for regularization. Finally, the embedded features are fed into a Fully Connected (FC) layer to predict delay until emergence for each country.

\subsubsection{Physics Informed GCN (PI-GCN)}

For comparison, we use a PINN-inspired~\cite{raissi2019physics} approach to make predictions based on dynamics for the Delay Prediction task. This model utilizes a 2-layer GCN architecture, incorporating dropout and two GraphNorm layers which are applied to the output of each GCN layer, along with Leaky ReLU activation functions.
Typically this GCN would be trained using a Mean Squared Error (MSE) loss, but we modify this loss function to steer our GCN towards dynamic predictions made by the dynamics-based model. 

The adjusted loss function is given by:
\begin{align}
    \text{MSE}_{\text{adjusted}} &= \frac{1-p}{N} \sum_{i=1}^{N} (\hat{y}_i - y_i)^2 + \frac{p}{N} \sum_{i=1}^{N} (\hat{y}_i - \dot{y}_i)^2 
\end{align}
where N is the number of countries, $\hat{y}_i$ is our model's output, $y_i$ is the ground truth, and $p$ is a hyperparameter ranging from 0 to 1, signifying the influence of $\dot{y}_i$, the output from the dynamics-based model. This modification ensures that, during training, the model is penalized more when its prediction deviates from the linear model's prediction, providing a form of dynamics-informed regularization. The best $p$ value was found to be 0.1 by grid search.

\subsection{Training Procedure} \label{ssec:train}

\begin{algorithm}
\caption{Retrospective Training and Validation}
\label{alg:training_validation}
\begin{algorithmic}[1]
\FOR{$variant$ in variants}
    \STATE dates $\gets$ unique dates in which $variant$ exists
    

    \STATE $model\_initial\_weights$ $\gets$ random

    \FOR{$d$ in dates}
        \STATE $retro\_D$ $\gets$ all data before $d$
        \STATE $processed\_D$ $\gets$ pre-process($retro\_D$)
        \STATE $dataset$ $\gets$ structure $processed\_data$ into graphs
        \STATE $train\_D, val\_D$ $\gets$ $dataset$ split 80/20\% 

        \STATE $model$ $\gets$ initialize with $model\_initial\_weights$
        \STATE $epochs$ $\gets$ 100
        \STATE $early\_St$ $\gets$ Early Stopper(patience=3)
        \STATE $model$.train($epochs$, $early\_St$, $train\_D$, $val\_D$)
        \STATE $best\_model\_weights$ $\gets early\_St$
        \STATE $model\_initial\_weights$ $\gets$ $best\_model\_weights$
    \ENDFOR
\ENDFOR
\end{algorithmic}
\end{algorithm}

All training and validation of the models were performed retrospectively, as outlined in Algorithm \ref{alg:training_validation}. This means that the model is trained and validated at each time step using only the ''observed" data available up to that time step, denoted as $d$. For each variant, the retrospective algorithm iterates through all the biweekly data associated with that variant, training the model only on the weeks that have already passed (i.e., the observed data). Note that the pre-processing is also done retrospectively, accounting for any smoothing, calculation of the $S$ values, or interpolation in the case of the dynamics-based model. This ensures that the approach is applicable in a prospective setting, where no data from the future is available.

The train/validation split is also done temporally to ensure that validation is performed only on the most recently observed data. This aims to achieve a better fit to the latest data, enhancing the model's predictive performance. An Early Stopper monitors the validation loss and halts the training when overfitting is detected, signified by a constant and substantial increase in the validation loss. The Early Stopper saves the best-performing weights, which are reused in the next iteration. This practice eliminates the need for the model to start training from scratch, leveraging knowledge gained from previous iterations.

We use MSE, or its adjusted version, as the loss function. Another crucial aspect is how to handle data related to variants that either do not appear in a given country or appear but fail to become dominant, i.e. rapidly diminish. In such cases, where regression targets would be undefined (potentially infinite), we address this challenge by creating a mask. This mask is employed to conceal nodes corresponding to these specific variants/country pairs, ensuring they do not contribute to our training loss. Essentially, our graph models are not trained on these nodes.

\section{Experiments}\label{sec:exp}

\subsection{Dataset} \label{ssec:data}

To validate our methodology, we've used a combination of multiple datasets which are detailed below.

\subsubsection{Covariant Data}
CoVariants \cite{covariants} is our primary dataset which provides the current state of variants and mutations of interest for SARS-CoV-2. This data is enabled by GISAID \cite{gisaid} and ranges from August 2020 through October 2023, with a bi-weekly resolution for 36 variants starting from 20A.EU1 to 23F.Omicron, which is current as of this work. CoVariants follows the NextStrain Clade schema \cite{nextstrain-clade}, in which variants can descend from other variants. 

\subsubsection{Border Restrictions Data}
We considered the border restrictions data provided in the OxCGRT \cite{hale2021global} dataset, with details on what level of border controls were enacted by governments in real-time from Jan 1, 2020, to Jan 8, 2023. OxCGRT measured the variations in government responses using their COVID-19 Government Response Stringency Index, which is a simple additive score of nine indicators ranging from school closures to vaccination policies. Of these indicators, we picked international travel closure controls since they best represent the goal of this paper in capturing the emergence and cross-border transmission of variants. These values are encoded in an ordinal scale as follows: 0 - no measures, 1 - screening, 2 - quarantine on high risk region, 3 - ban on high risk regions, 4 - total border closure. The recorded values were then scaled down to a range between 0.1 and 1, with 1 indicating no border restrictions and 0.1 representing complete border closure. Note that 0.1 was chosen over 0, acknowledging the possibility of some mobility between connecting countries due to necessary trade or border crossings via open land routes. Furthermore, given that our CoVariants data extends beyond January 2023, we extend OxCGRT by making the assumption that all border restrictions have remained unchanged since Jan 8, 2023.

\subsubsection{Country Graphs} 
We considered two graphs representing the interconnection between countries. This dataset depicts country neighborhoods according to the Correlates of War (CoW) database. Countries were also deemed neighbors if they share a land or river border or are located within 24 miles of each other across bodies of water. 
\textbf{Adjacency:} We used the country adjacency dataset \cite{countryAdjacencyData}, which offers a list of countries and their neighboring countries.
\textbf{OpenFlights:} This is a collection of regular flights along with their routes mapped from airports around the world \cite{OpenFlights2023}. Given the scope of the current work, we primarily focus on airports and the flight route data, and was mainly used during ablation studies. Airports data is a UTF-8 encoded collection of over 14,000 airports, train stations and ferry terminals across the world as of January 2017. It consists of information like city/country, IATA codes, geographic coordinates and the type of the airport (air terminal, ferry station, etc.). Routes data provides information about 67663 directional routes between 3321 airports on 548 airlines. This dataset was used in an ablation study to validate the effectiveness of utilizing these flight routes as edges in our graph creation process.

A specific adjustment in preprocessing was necessary
for the OpenFlights dataset, involving the mapping of city
routes to their respective country routes. Leveraging the provided airport data, which contains comprehensive city and
country details, facilitated this mapping process. The transformation ensured that subsequent analyses operated consistently at the country level.

\subsection{Benchmark Pipeline} \label{ssec:bench}

\begin{figure}[!ht]
\centering
 \includegraphics[width=\linewidth]{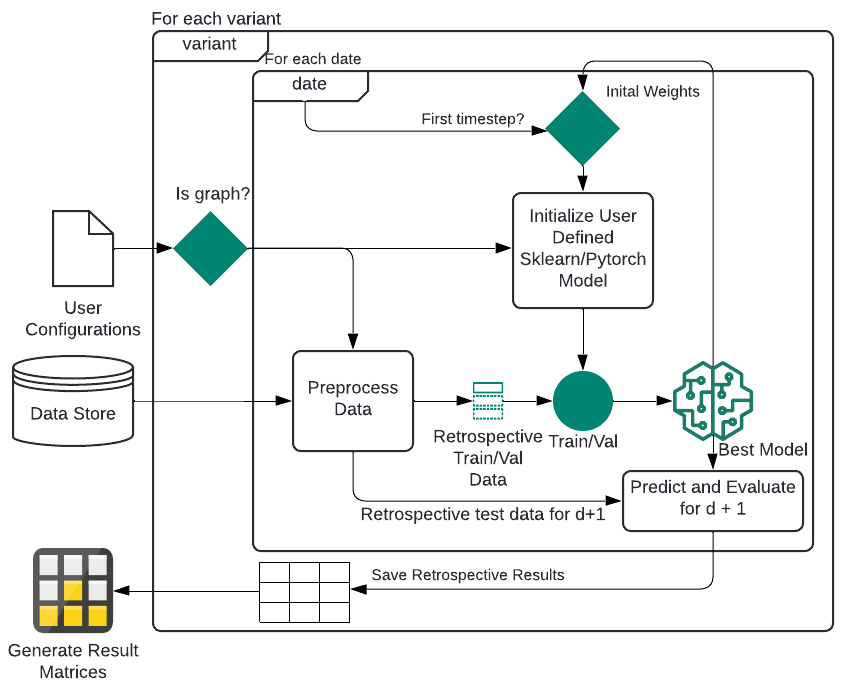}
 \caption{\small Benchmarking Pipeline}
 \label{fig:Benchmark}
\end{figure}

Since we are the first to attempt solving this problem, we provide a benchmarking tool for the community. Figure \ref{fig:Benchmark} presents the comprehensive pipeline employed for retrospective training, validation, and testing of diverse models. Its design prioritizes user-friendliness, allowing seamless integration of new PyTorch models or direct utilization of scikit-learn models through a configuration file. Users only need to specify whether the model requires graph data, leaving the pipeline to handle the rest. Additionally, the data pre-processing is encapsulated in a separate function, providing users with the flexibility to make modifications. 

As shown in Figure \ref{fig:Benchmark}, the evaluation is also being done retrospectively. It means that for each date that the model is trained on, it is used to predict for the subsequent date representing a 1-timestep ahead forecast. This aligns with the biweekly nature of the data, resulting in a two-week ahead forecast. Furthermore, model performance is being evaluated per variant, meaning that for each given variant we retrospectively assess the trained models over time. 

In the context of delay prediction, considering the biweekly data, the target is consistently a multiple of 14 days. Therefore, the output of our models is always rounded up to the nearest multiple of 14. Furthermore, evaluations exclude countries where a variant has already become dominant or never achieves dominance.

\subsection{Pre-processing}

The initial pre-processing stage aligned the country names across diverse datasets, with an emphasis on inclusion based on data availability. Only countries present in all datasets were retained, ensuring a standardized set for subsequent analyses. 
For each country, the prevalence of each variant was calculated at each timestep. Variants with a prevalence of less than 5\% were disregarded in our analysis. To calculate the prevalence ratios, the two variants with the highest prevalence were identified for each timestep. For each variant on that particular day, the prevalence ratio was calculated. Finally, variants that did not reach a prevalence ratio of 0.2 or those that were present in less than three countries were filtered out. These steps ensured that only variants with a significant presence and wide geographic distribution were included in the subsequent analysis.
All subsequent pre-processing steps were conducted retrospectively. This implies that as time progresses, more data is used for computing the global growth rates ($S$) and for training.

\subsection{Benchmark Evaluation Metrics}
To measure performance, we define Mean/Median Absolute Errors (MAE/MedAE) as follows:
\begin{align}
    \text{MAE}_{t, v} &= \frac{1}{N_s} \sum_{i=1}^{N_s} | \hat{y}_{i, t, v} - y_{i, t, v} | \\
    \text{MedAE}_{t, v} &= \text{median}\left(| \hat{y}_{i, t, v} - y_{i, t, v} |\right) 
\end{align}

where $i$ ranges from 1 to $N_s$ and $N_s$ represents the number of valid countries to predict for. The $t$, $v$ subscripts indicate that these errors are calculated for each variant at each timestep. Since this problem is retrospective, we also evaluate performance over time by finding the mean and median of the errors:
\begin{align*}
    \text{MMAE}_v &= \frac{1}{T} \sum_{t=1}^{T} (\text{MAE}_{v, t}) \\
    \text{MMedAE}_v &= \frac{1}{T} \sum_{t=1}^{T} (\text{MedAE}_{v, t}) \\
    \text{MedMAE}_v &= \text{median}(\text{MAE}_{v, t}) \\
    \text{MedMedAE}_v &= \text{median}(\text{MedAE}_{v, t})
\end{align*}

where $t$ ranges from 1 to $T$ and $T$ represents the total number of timesteps for which the variant $v$ circulates before reaching total dominance, i.e. it globally reached all its targets and begins to disappear. Finally, we conduct a mean across all the variants $v$ to get our four error metrics: MMAE, MMedAE, MedMAE, and MedMedAE. While all these metrics offer valuable insights, our main focus lies on MedMAE and MedMedAE, emphasizing median performance over time. This choice is made due to certain data inconsistencies, such as variants appearing in a single country for an extended period before spreading to other countries, resulting in large errors that may skew results.

\subsection{Benchmark Models} \label{ssec:benchModels}
We implemented a number of benchmark models that act as baselines for comparisons. 

\textbf{Mean Model -- Delay Prediction}: 
This model predicts delays based on average delays on the variants between the countries under consideration. Specifically, at time $t$, for a new variant $i$ on a target country $Y$, suppose $Q$ is the set of all countries where the variant has appeared. Then, the predicted time of arrival is given by
\begin{equation}
{t^{Y}_i = t + \mbox{median}_q\left(\mbox{mean}_j \left(\Delta_{j}(Y, q) \right)\right)\,.
}\end{equation}

Here $\Delta_j(Y, q)$ is the delay between the appearance of variant $j$ from country $q$ to country $Y$. We take the mean of $\Delta_j(Y, q)$ over all prior variants $j$ that appeared in both countries $Y$ and $q$. Finally, we take the median of the delays over all countries in $Q$ where variant $i$ has already appeared.

\textbf{Dynamics-based Model -- Delay Prediction}: This is the derived dynamics-based linear model discussed in the previous section. This model serves as a baseline mechanistic model, which, under the right assumptions, offers an explainable and efficient method for calculating delays.

\textbf{Decision Tree}: We use Decision Tree regression models to address Delay Prediction. This involves transforming the dataset into an $N \times 5$ matrix. Here, $N$ represents the total number of samples, where each sample corresponds to the 5 features of a given variant at a specific point in time in a specific country. Meaning $N = T \times V \times C$, where $T$ is the total number of timesteps, $V$ is the total number of variants, and $C$ is the total number of countries. These trees use the Gini impurity and mean squared error for their criterion functions. No maximum depth was set for either.

\textbf{Multi Layer Perceptron}: We employ a Multi-Layer Perceptron (MLP) regressor for Delay Prediction. They were configured with two hidden layers, consisting of 16 and 8 neurons, respectively. The selection of hyperparameters, including the number of hidden layers, activation function, and solver, was determined through comprehensive experimentation and optimization tailored to our specific problem. The training procedure closely followed the approach outlined in Section 3.3, with the only distinction being the structuring of our dataset into an $N \times 5$ matrix, as described for the decision tree. A batch size equivalent to the number of countries $C$ was used, where each batch represented the list of countries and their features at each timestep for each variant, ensuring that the gradient updates were performed collectively for each group of variants at a given timestep.

\textbf{Gated Recurrent Unit Network}
We utilize a simple GRU network to encode the temporal information of variant transmissions and predict theirdelay. To train this model, first, we feed temporal features of the variant to the GRU architecture which embeds them for each country distinctly. Then, we flatten the embedded features and append them with $S$ and directly give them to an FC layer for predicting the time it takes for that variant to appear in a specific country. We employ GraphNorm layer here after the concatenation to accelerate training by smoothing graph aggregation distributions.

\textbf{Graph Convolutional Network (GCN)}
We consider a simple GCN, with a basic architecture in an attempt to capture local relationships and features within the graph structure. It is comprised of two GCN layers, which output 32 and 16 channels respectively. Dropout is strategically employed between these two layers for regularization. The final graph embedding is flattened and fed into an FC layer which outputs the delay for each country.

\textbf{Encoded Dynamics GCN}
Each country has a variable number of circulating variants, each with different growth rates $S$ (as derived earlier). These growth rates can be concatenated and represented as a feature vector $S_{qt}$, which correspond to the variants present in a region $q$ at time $t$. We can learn a latent vector representation $Z_{qt}$ of this data through the use of encoders. We implement this in two ways:

\textbf{Using an Autoencoder (AE-GCN):} We employ a simple autoencoder with two linear layers and ReLU activation in the encoder and decoder blocks. The model takes the feature vector $S_{qt}$ as input. We train the autoencoder using MSE loss to reconstruct the feature vector $S_{qt}$. After training, we take the intermediate latent encoding $Z_{qt}$ output from the encoder and append it to the feature matrix before constructing a spatio-temporal snapshot graph data for the GCN model.

\textbf{Embedded Encoder (EE-GCN):} Using an autoencoder implies that the model is guided towards learning an embedding using the decoder output, and these two are inherently separate models. The encoder weights are also unaffected during graph training. An alternative approach is to embed this into the larger GCN model. In this method, as before, the variable-sized $S_{qt}$ is fed to an encoder block (consisting of two linear layers with ReLU activation). Instead of using a decoder, we directly concatenate the resulting latent vector representation $Z_{qt}$ to the node features fed into the GCN. This means the encoder weights are updated in the graph training loop, allowing the model to learn a better latent representation.

\subsection{Results}\label{sec:metrics}

In this section, we present our key results for Emergence Delay Prediction. We previously noted that prior research has concentrated on predicting the prevalence of a variant after its arrival in a specific region, with logistic regression demonstrating effectiveness in estimating its prevalence over time. However, as logistic regression can manage this task relatively easily, our evaluation focused solely on forecasting a variant's prevalence before its arrival in a specific region. In other words, during evaluation, we excluded cases where a variant is present in a region but has not yet reached $\theta$.

Numerous models were tested and configured as benchmarks to test against the performance of our proposed method. Other implementations include a trivial \textit{Mean Model}, the baseline \textit{Dynamics-based Model}, a Decision Tree, a Multilayer Perceptron (MLP), a GRU, and a GCN without an adjusted loss function. In brief, the Mean Model predicts delays based on the average delays of all prior variants between all the countries in which the variant appeared and our intended country. The Decision Tree, MLP, and GRU regression models make use of $T$ extra features which consist of the average $\vec{r}_{c,i}$ of neighboring countries transforming the dataset into a matrix of size $N\times(2T+1)$. Here, $N = D \times V \times C$ is the total number of samples, $D$ is the total number of retrospective dates, $V$ is the total number of variants, and $C$ is the total number of countries. The full details on these models can be found in the Appendix.

To measure the performance of the models, we use the defined Mean/Median Absolute Errors (MAE/MedAE) in the Appendix and measure the performance over time by finding the mean and median of these errors as:
\begin{align*}
    \text{MedMAE}_v = \text{median}(\text{MAE}_{v, t}) \mbox{,  } \\
    \text{MedMedAE}_v = \text{median}(\text{MedAE}_{v, t}),
\end{align*}
where $t$ ranges from 1 to $T_D$, $T_D$ being the total number of timesteps for which variant $v$ circulates before reaching total dominance, i.e. it globally reaches all its targets and begins to disappear. Finally, the mean across all the variants $v$ gets our two error metrics: MedMAE, and MedMedAE. 

\begin{table}[!ht]
    \centering
     \caption{Prediction Errors}
 \label{tab:full_experiment_results_R}
    \begin{tabular}{|c|c|c|}
    \hline
        \textbf{Model} & \textbf{MedMedAE} & \textbf{MedMAE} \\ \hline
        Mean Model & 3.2 & 3.87  \\ \hline
        Dynamics-based Model & 1.86 & 2.9  \\ \hline
        Decision Tree & 3.48 & 3.43  \\ \hline
        MLP & 2.38 & 2.48  \\ \hline
        GRU & 1.5 & 1.37  \\ \hline
        GCN & 1.32 & 1.43  \\ \hline
        \textbf{FA-GCN} & \textbf{1.27} & \textbf{1.29}  \\ \hline
        PI-GCN & 2.24 & 2.01  \\ \hline
        AE-GCN & 1.82 & 2.25 \\ \hline
        EE-GCN & 2.00 & 2.49 \\ \hline
    \end{tabular}
\end{table}

Table~\ref{tab:full_experiment_results_R} presents the models' performance metrics for Delay Prediction. The results indicate that our proposed FA-GCN model outperforms all other models in terms of MedMedAE and MedMAE, showcasing its effectiveness in capturing temporal dependencies for Delay Prediction. The dynamics-based model also has a significantly worse MedMAE than MedMedAE. This suggests that the dynamics-based model is susceptible to outlier predictions for some country pairs, which skews its results. Additionally, we observe that the PI-GCN does not surpass its GCN counterpart, highlighting the unnecessary use of dynamics as a regularization of the loss when employing the correct features. 

\begin{figure*}[!ht]
     \centering
     \includegraphics[width=0.7\linewidth]{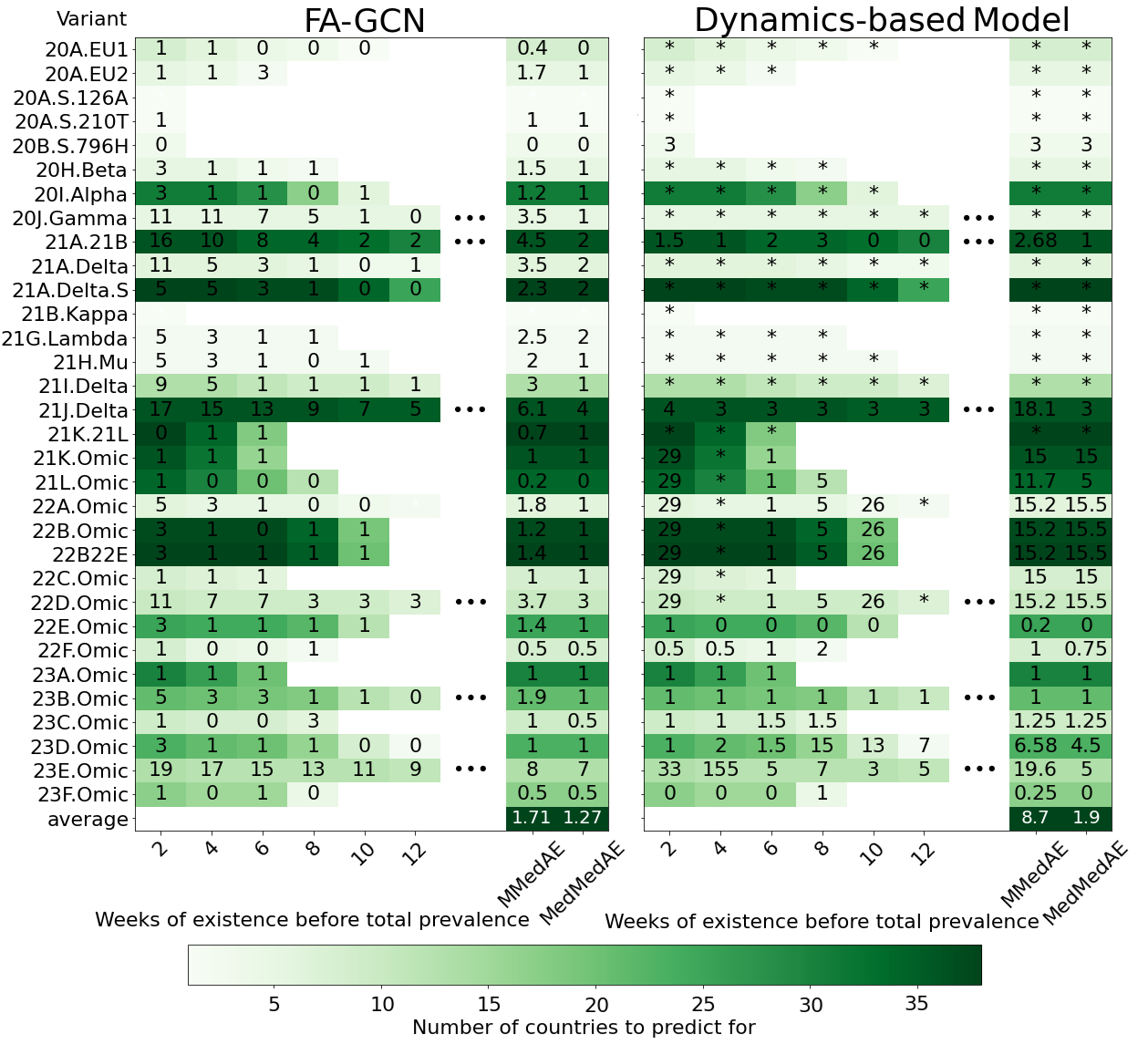}
     \caption{\small The heatmaps display both the errors and the number of countries being predicted for each variant at each timestep until they are no longer dominant anywhere.}
      \label{fig:Overall Figure}
\end{figure*}

\section{Discussion}

\subsection{Results Analysis}

Figure \ref{fig:Overall Figure} provides a detailed representation of the results for the baseline dynamics-based model and the FA-GCN model. The heatmaps illustrate the number of countries eligible for prediction at each timestamp. The numbers indicate the errors over time for each variant throughout their existence. Notably, several variants in the dynamics-based model are denoted by $*$, signifying instances of model or evaluation failure. These failures stem from three primary factors.  Firstly, the linear model requires a minimum of two common variants between two countries to build an effective model. Secondly, the linear model can encounter significant challenges when variant $j$ is identical in growth rate to variant $k$ (i.e., $S_j = S_k$). In such cases, the linear model fits a plane solely along a fixed axis and will thus always predict some constant $\frac{C}{S_{ij}}$, as seen in Equation \ref{eq:final}. But as time passes and variant $j$ is no longer identical to variant $k$ for another variant, this model fails drastically since it would still fit on a fixed axis. These scenarios were more prevalent in the early stages of the pandemic when only a few variants were circulating, leading to their exclusion from the dynamics-based model's training process. To ensure fair comparisons, all final errors across all models were calculated excluding these variants. Finally, there are cases where a variant appears in a country but fails to reach the threshold $\theta$. These instances were disregarded during the evaluation of all models as they were considered trivial.

The observations from Figure \ref{fig:Overall Figure} give us two insights. Firstly, the primary source of error is associated with variants that persist for an extended duration, such as 21J.Delta or 23E.Omicron. Examining the color gradients, these variants appear to linger in only a few countries before spreading to others. In reality, variants typically do not endure for such prolonged periods without either being superseded by another variant or transmitting to additional countries \cite{Chen2023,Verywell_Health_2023}. This suggests a potential issue in the data capturing these lingering variants. Secondly, despite this being a major source of error for FA-GCN, it exhibits overall robustness to outlier predictions, unlike the dynamics-based model as seen in Figure \ref{fig:Overall Figure}. This, in turn, severely affects the results of the PI-GCN model. 

\begin{figure*}[!ht]
    \centering
    \begin{subfigure}[b]{0.35\textwidth}
        \includegraphics[width=\textwidth]{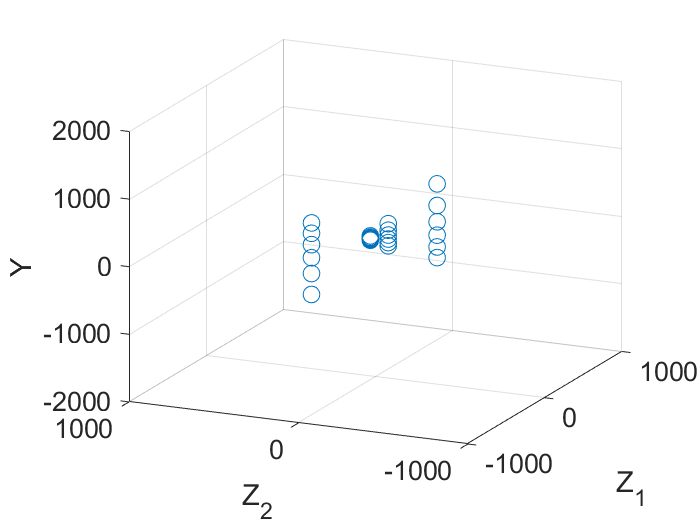}
        \caption{Good fit}
        \label{fig:goodfit}
    \end{subfigure}
    \begin{subfigure}[b]{0.35\textwidth}
        \includegraphics[width=\textwidth]{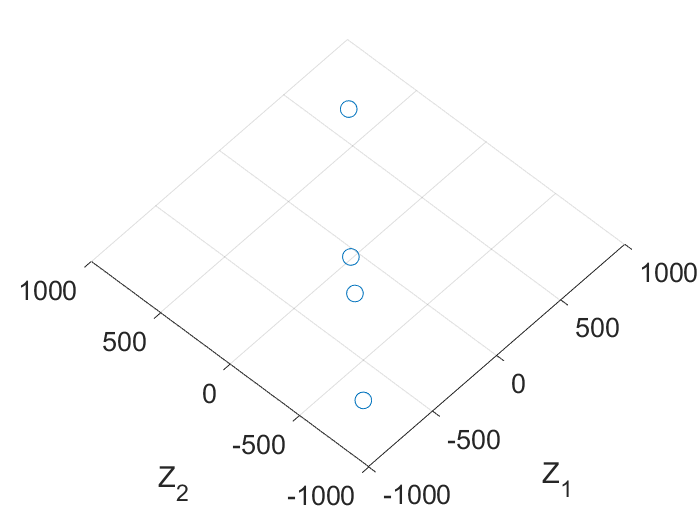}
        \caption{Good fit (2D representation)}
        \label{fig:goodfit1}
    \end{subfigure}
    \begin{subfigure}[b]{0.35\textwidth}
        \includegraphics[width=\textwidth]{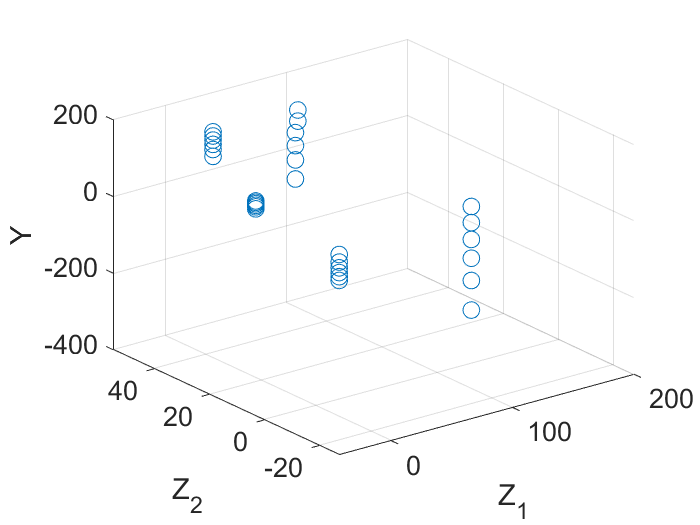}
        \caption{Bad fit}
        \label{fig:badfit}
    \end{subfigure}
    \begin{subfigure}[b]{0.35\textwidth}
        \includegraphics[width=\textwidth]{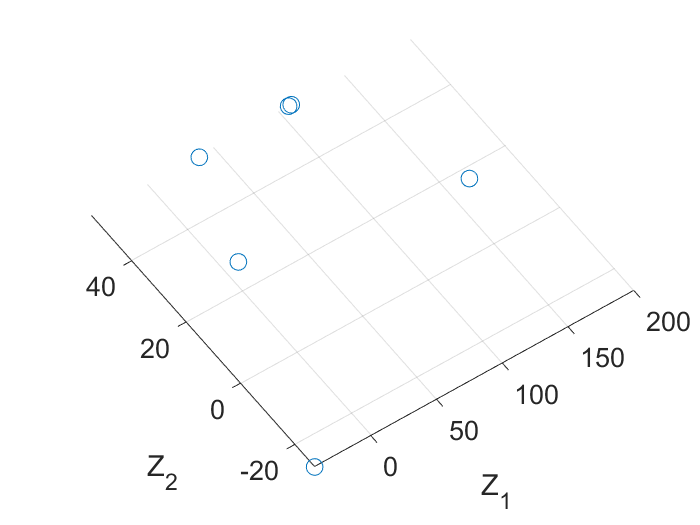}
        \caption{Bad fit (2D representation)}
        \label{fig:goodfit2}
    \end{subfigure}
    \caption{3D plots of the X and Y points in space, from equation \ref{eq:LM}, and their 2D representations. Examples are taken for two linear models fitting on data for 22A Omicron.}
    \label{fig:good_bad_fits}
\end{figure*}

While the dynamics-based model does output good predictions, it fails often and dramatically in some cases. Figure \ref{fig:good_bad_fits} presents examples of 3D points in space utilized for fitting linear models. In Figure \ref{fig:goodfit}, the 3D points form a space where a plane can be fitted exceptionally well. However, as newer COVID variants emerge, the dynamics may evolve, causing some initial assumptions, such as full cross-immunity \cite{Chen2023}, to become invalid. This evolution makes it more challenging to perfectly fit a linear model, as depicted in Figure \ref{fig:badfit}.

\subsection{Ablation Study}

\subsubsection{Dynamics}

\begin{table}
\centering
\caption{Feature Ablations MedMedAE}
\label{tab:dynamic_ablations_MAE}
\begin{tabular}{c|c|c|c|c|c}
\hline
\textbf{Model}  & \textbf{$[r|S]$}    & \textbf{$r$} & \textbf{$[p|S]$} & \textbf{$p$} & \textbf{$S$} \\ \hline
Decision Tree & 3.48 & 3.0 & \textbf{2.81} & 3.0 & 2.86 \\ 
MLP & \textbf{2.38} & 2.67 & 2.69 & 2.55 & 2.76 \\
GRU & \textbf{1.50} & 1.50 & 1.95 & 2.32 & n/a \\
GCN & \textbf{1.32} & 1.50 & 1.78 & 2.32 & 2.38 \\
FA-GCN & \textbf{1.27} & 1.55 & 1.73 & 1.73 & n/a \\
\end{tabular}
\end{table}

We examine the usefulness of integrating the extracted dynamics as inputs for our models. In this analysis, we train the same models using different sets of features. Specifically, we investigate the effectiveness of using $S$ as a feature and the impact of using a time series of $r$. Where $r$ excludes $S$ as a feature, relying solely on the time series of prevalence ratios rather than the prevalence $p$. Additionally, we investigate the effectiveness of $S$ by comparing $[p|S]$ to $p$ and utilizing $S$ as the only feature.

From the results in Table \ref{tab:dynamic_ablations_MAE}, we observe that the $[r|S]$ combination outperformed all other configurations except for the case of the Decision Tree. There is also an improvement when utilizing $r$ over $p$ for the temporal-based models, with the GCN performing similarly and the others exhibiting poorer performance. This suggests that the temporal models effectively encode the latent information contained in the time series of $r$, while the simpler models struggle to do so. Additionally, inclusion of $S$ as a feature enhances results across all cases. Combining $S$ and $r$ significantly improves performance in most cases, indicating a synergistic effect likely stemming from the linear relationship between the two, as previously discussed. It is important to note that the GRU and FA-GCN were not trained using just $S$ since excluding the time series encoding reduces these models to an MLP and GCN, respectively. 

\subsubsection{Graphs}
\begin{table}
\centering
\caption{Graph Ablation Results}
\label{tab:graph_ablations}
\begin{tabular}{|c|c|c|}
\hline
\textbf{Model}  & \textbf{MedMedAE}    & \textbf{MedMAE} \\ \hline
\textbf{FA-GCN} & \textbf{1.27} & \textbf{1.29} \\ \hline
FA-GCN\_30 & 1.46 & 1.45 \\ \hline
FA-GCN\_60 & 1.57 & 1.46 \\ \hline
FA-GCN\_flights\_adj & 1.95& 1.46 \\ \hline
FA-GCN\_flights & 1.88 & 1.53 \\ \hline
FA-GCN\_fullcon & 2.10 & 1.67 \\ \hline
FA-GCN\_WoEW & 1.46 & 1.46 \\ \hline
\end{tabular}
\end{table}

When designing the structure of our graph, several considerations were taken. Firstly, we acknowledged that certain nodes (countries) might be more susceptible to noise due to reporting inconsistencies or a lack of reporting, leading to unreliable data. To address this issue, we explored the option of using only countries with consistently reported data, focusing on the top 30 (FA-GCN\_30) and top 60 (FA-GCN\_60) countries out of the total 87 (FA-GCN) countries in our dataset. 

Additionally, we examined different approaches for creating edges in our graph. While country adjacency served as the base (FA-GCN), we also explored the option of connecting countries through flight routes using the OpenFlights dataset. This exploration included a combination of country adjacency and flight routes (FA-GCN\_flights\_adj), as well as utilizing only flight routes (FA-GCN\_flights). 

To ensure a thorough comparison, we also tested a fully connected graph (FA-GCN\_fullcon), where every country was connected to every other country. This configuration was expected to test the limits of connectivity by removing the specificity of relationships, allowing us to see if more generalized connections might yield better results. Additionally, we included a model without edge weights (FA-GCN\_WoEW) to serve as a baseline, assessing the importance of dynamic weights in the performance of the graph model.

From Table \ref{tab:graph_ablations}, it is evident that our selected graph design, based on country adjacency, outperforms alternative configurations. The number of nodes (countries) in the graph does not significantly impact the model performance, showcasing its robustness. In contrast, a fully connected graph performs the worst, indicating that simply increasing the number of connections without considering the relevance of those connections can introduce noise and degrade the model's accuracy. It emphasizes the importance of meaningful relationships between nodes, rather than indiscriminately connecting all nodes.

Interestingly, the use of flight routes generally results in increased errors when compared to the adjacency-based model. While this might seem counter-intuitive, considering that flights could connect distant countries and continents. It's crucial to note that, based on the definition of country adjacency used by the dataset, countries like the USA and Russia are considered adjacent. As a result, they act as a link between continents. In such cases, the adjacency-based connections might already encapsulate the critical pathways for information flow, making the additional flight route data redundant or even misleading. Moreover, the incorporation of dynamic weights appears to improve model performance, as seen when comparing FA-GCN with FA-GCN\_WoEW.

Overall, these results underline the importance of carefully considering the graph structure and the nature of the connections when designing GCN models. While flight routes might seem like a logical addition, their effectiveness depends on how well they complement the existing adjacency information. In this case, it appears that the adjacency-based graph design captures the essential relationships better than the other configurations.

\subsection{Implementation}
The pipeline depicted in Figure \ref{fig:Benchmark}, was implemented in Python. The implementation was carried out on a machine equipped with a 32-core Intel(R) Xeon(R) Gold 5218 CPU running at 2.3 GHz and 64 GB of RAM. Focusing on the example of variant 23F.Omicron at a single timestep, which contains the majority of the data, and the heaviest model (FA-GCN), the average runtimes for different components were approximately 32 seconds for data pre-processing, 20 seconds for training, and 0.01 seconds for inference. The entire retrospective pipeline takes approximately 2 hours to complete. Notably, the most time-consuming aspect of the pipeline is the data pre-processing step, which could potentially be optimized further through memorization.

\subsection{Limitations}
One limitation of our approach in addressing the problem of Delay Prediction lies in handling situations where a variant might not even become dominant in a given country. To tackle this issue, we define the following problem:

\begin{problem}[Dominance Prediction]
    Given the prevalence (proportion) of a variant in regions $A_1, A_2, \dots$, predict if the variant will ever reach a proportion $\theta$ in region $B$.
\end{problem}

In future work, we plan to address this issue by adopting a two-step approach. First, we solve Dominance Prediction, which aims to predict whether a variant will become dominant in a given region. If the variant is predicted to become dominant, we can then proceed to solve Delay Prediction.

\section{Conclusion} 
We addressed the challenges of predicting variant delay across countries. The derivation of variant dynamics provided a theoretical foundation, which was used to engineer relevant features as well as a novel baseline dynamics-based model. We demonstrated that our dynamics feature-augmented GNN approach outperformed all other methods. Through comprehensive experiments and analysis, we demonstrated the effectiveness of our design choices, providing valuable tools for understanding and predicting the intricate relationships and connectivity patterns between nations. Furthermore, as we are the first to address the proposed problem, we provided a comprehensive benchmark and made our full pipeline available in an effort to facilitate research in the field.

\section*{Acknowledgements}
This work was supported by the Centers for Disease Control and Prevention and the National Science Foundation under the awards no. 2135784, 2223933, and 2333494.

\bibliographystyle{ieeetr}
\bibliography{refs}

\end{document}